\documentclass[reprint,numerical]{revtex4-1}
\usepackage{srctex}
\usepackage{amsmath}
\usepackage{amsfonts}
\usepackage{amssymb}
\usepackage{epsfig}
\usepackage{graphicx}
\usepackage[dvipsnames]{xcolor}
\newcommand{\ve}{\varepsilon}

\begin{document}

\title{Thermodynamics of low-dimensional trapped Fermi gases}

\author{Francisco J. Sevilla}
\email{fjsevilla@fisica.unam.mx}
\affiliation{Instituto de F\'{\i}sica, Universidad Nacional Aut\'onoma de M\'exico,\\ Apdo. Postal 20-364,
01000 M\'exico D.F., MEXICO}

\begin{abstract}
The effects of low dimensionality on the thermodynamics of a Fermi gas trapped by isotropic power law potentials are analyzed. Particular attention is given to different characteristic temperatures that emerge, at low dimensionality, in the thermodynamic functions of state and in the thermodynamic susceptibilities (isothermal compressibility and specific heat). An \emph{energy-entropy} argument that physically favors the relevance of one of these characteristic temperatures, namely, the non vanishing temperature at which the chemical potential reaches the Fermi energy value, is presented. Such an argument allows to interpret the nonmonotonic dependence of the chemical potential on temperature, as an indicator of the appearance of a thermodynamic regime, where the equilibrium states of a trapped Fermi gas are characterized by larger fluctuations in energy and particle density as is revealed in the corresponding thermodynamics susceptibilities.
\end{abstract}

\maketitle

\section{\label{SectI} Introduction}
The discovery of the quantum statistics that incorporate Pauli's exclusion principle \cite{Pauli1925}, made independently by Fermi \cite{Fermi1926} and Dirac \cite{Dirac1926}, allowed the qualitative understanding of several physical phenomena---in a wide range of values of the particle density, from astrophysical scales to sub-nuclear ones--- in terms of the ideal Fermi gas (IFG). 
The success of the explicative scope of the ideal Fermi gas model relies on Landau's Fermi liquid theory where, fermions interacting repulsively through a short range forces can be described in some degree as an IFG. The situations changes dramatically in low dimensions, since Fermi systems are inherently unstable towards any finite interaction \cite{SolyomAdvPhys1979,VoitRPP1995,GuanRMP2013}, thus the IFG in low dimensions becomes an interesting solvable model to study the thermodynamics of possible singular behavior.

On the other hand, the experimental realization of quantum degeneracy in trapped atomic Fermi gases \cite{DemarcoSc,TruscottScience01,GranadePRL02,HadzibabicPRL03,FukuhuraPRL07} triggered a renewed interest, over the last fifteen years, in the study not only of interacting fermion systems \cite{ChinScience04,CampbellScience09,ZwierleinNature06,ShinPRL08} but also of trapped ideal ones as well  \cite{ButtsPRA97,Schneider98,LiPRA98,Thilagam98,VignoloPRL00,BrackPRL00,BruunPRA00,GleisbergPRA00,TranPRE01,AkdenizPRA02,GretherEPJD2003-2,VignoloPRA03,AnghelJPhysA03,TranJPhysA03,vanZylPRA03,Mueller04,AnghelJPhysA05,SongPRA06,FarukaAPhysPolB2015}. Indeed, the nearly ideal situation has been experimentally realized by taking advantage of the suppression of $s$-wave scattering in spin-polarized fermion gases due to Pauli exclusion principle and of the negligible effects of $p$-wave scattering for the temperature ranges involved. Further, the control achieved on the experimental settings has open the possibility to directly test a variety of quantum effects such as Pauli blocking \cite{DemarcoPRL}, and to design experiments to probe condensed matter models, though much lower temperatures are needed to achieve the phenomena of interest. On this trend, experimentally new techniques are being devised to cool further a cloud of atomic fermions \cite{BernierPRA09,CataniPRL09,StamperPhysics2009,BernierPRA09,PaivaPRL2010}. Techniques based on the giving-away of entropy by changing the shape of the trapping potential has resulted of great importance and, as in many instances, a complete understanding of trapped non-interacting fermionic atoms would result of great value.

In distinction with the ideal Bose gas (IBG), which suffers the so-called Bose-Einstein condensation (BEC) in three dimensions, the IFG shows a smooth thermodynamic behavior as function of the particle density and temperature, this however, does not precludes interesting behavior as has been pointed out in Refs. \cite{AnghelJPhysA03,Romero2011}, where it is suggested that the IFG can suffer a condensation-like process at a characteristic temperature $T^{0}$. Arguments based on a thermodynamic approach in support of this phenomenon are presented in Ref. \cite{Romero2011}, where the author suggest that the change of sign of the chemical potential, which defines the characteristic temperature $T^{0},$ marks the appearance of the condensed phase when the gas is cooled.  

Truly, the significance of $\mu$ has motivated the discussion of its meaning and/or importance at different levels and contexts \cite{cook_ajp95,baierlein,job2006,MuganEJP2009,ShegelskiSSC86,LandsbergSST87,ShegelskiAJP2004,KaplanJStatPhys2006,PandaPramana2010,SevillaEJP2012,SevillaArxiv2014,SalasJLTP2014}.
For the widely discussed---textbook--- case, namely the three-dimensional IFG confined by a impenetrable box potential, the chemical potential results to be a monotonic decreasing function of the temperature, diminishing from the \emph{Fermi energy}, $E_{F}$, at zero temperature, to the values of the ideal classical gas for temperatures much larger than $k_{B}^{-1}(\hbar^{2}/m\lambda_{T}^{2})$, where $k_{B}$ is the Boltzmann's constant, $\hbar$ is the Planck's constant divided by $2\pi$, $m$ the mass of the particle and $\lambda_{T}=\sqrt{2\pi\hbar^{2}/mk_{B}T}$ is the thermal wavelength of de Broglie, where $T$ denotes the system's absolute temperature. A clear, qualitative, physical argument of this behavior is presented by Cook and Dickerson in Ref. \cite{cook_ajp95}. In comparison, the chemical potential of the IBG vanishes below a characteristic temperature, called the critical temperature of BEC, $T_{c},$ and  decreases monotonically for larger temperatures converging asymptotically to the values of the classical ideal gas. 

This picture changes dramatically as the dimensionality of the system $d$, is lowered. In two dimensions the IBG shows no off-diagonal-long-range order at any finite temperature \cite{HohenbergPR1967} and therefore the BEC transition does not occur. At this quirky dimension, the chemical potential of both, the Fermi and Bose ideal gases, decreases monotonically with temperature essentially in the same functional way \cite{leePRE97}, being different only by an additive constant, expressly, the Fermi energy. This results in the same temperature dependence of their respective specific heats at constant volume $C_{\mathcal{V}}$ \cite{mayPR64,leePRE97,anghelJPA02}. In general, this last outstanding feature occurs whenever the number of energy levels per energy interval is uniform as in the case of a one dimensional gas in an harmonic trap \cite{SchonhammerAJP96,CrescimannoPRA01}, or the case $s=d$ where $s$ is the exponent of the single-particle energy spectrum of the form $\varepsilon\propto p^{s},$ $p$ being the particle momentum \cite{PathriaPRE98}. 

In one dimension, the chemical potential of the IBG decreases monotonically with temperature, and as in the two dimensional case, this behavior is related to the impossibility of BEC as shown by Hohenberg \cite{HohenbergPR1967}, at finite, non-zero, temperature. In contrast, the chemical potential of the IFG exhibits a \emph{nonmonotonic} behavior: starts rising quadratically with $T$ above the Fermi energy instead of decreasing from it, and returns to its usual monotonic-decreasing behavior at temperatures that can be as large as twice the Fermi temperature (see Fig. 1 below, see also Fig. 1 in Ref. \cite{GretherEPJD2003}). This \emph{unexpected}, and not well understood behavior, can be exhibited mathematically by the Sommerfeld expansion \cite{cetina77,GretherEPJD2003} or by other methods \cite{LeeJMathPhys95,gretherIJMPB09,ChavezPhysicaE2011}, though no intuitive physical explanation of it, that predicts its appearance in the more general case, seems to have been given before \cite{footnote00}. This forms the basis for the motivation of the present paper.

After this excursus, one may conceive dimension two as a \emph{crossover} value for which the thermodynamic properties of ideal quantum gases are conspicuously distinct for $d>2$ than those for $d<2$. This can be seen in the specific heat, which in the case of the IFG exhibits a no-bump $\rightarrow$ bump transition as dimension is varied from 3 to 1 \cite{GretherEPJD2003} analogous to the well known cusp $\rightarrow$ no-cusp transition of the IBG specific heat. In the later case, the cusp marks the BEC phase transition while no physical meaning is yet given for the bump in the former case. 

In this paper we provide an analysis that attempts to explain the various features that are observed in the low-dimensional, trapped IFG, focusing in the nonmonotonic dependence on $T$ of the chemical potential. In section \ref{sect:2} the system under consideration is described, thermodynamics quantities are calculated and characteristics temperatures are introduced. In section \ref{sect:3} a heuristic explanation of the nonmonotonic dependence of the IFG chemical potential on temperature is given. In sections \ref{sect:4} and \ref{sect:5} the physical meaning of two relevant characteristic temperatures is given. Finally, conclusions and final remarks conform section \ref{sect:6}.

\section{\label{sect:2} General relations, calculation of the chemical potential and the thermodynamical susceptibilities}

We consider an IFG of $N$, conserved, spinless fermions in arbitrary dimension $d>0$. We assume a single-particle density of states (DOS) of the form \cite{PathriaPRE98,AnghelJPhysA03}
\begin{equation}\label{DOS}
g(\varepsilon)=G_{d,s}\, \varepsilon^{d/s-1},
\end{equation}
where $\varepsilon$ denotes the energy, $G_{d,s}$ and $s$ are positive constants, the former depends on $d$ and on the specific energy spectrum 
of the system, while the later is determined by the particular system dynamics.

Two instances lead to the power-law dependance in expression (\ref{DOS}): the first one is based on the
\textit{generalized energy-momentum relation} \cite{PathriaPRE98,AguileraEJP99} $\varepsilon_{k}=\mathcal{C}
_{s}k^{s}$, $k$ being the magnitude of the particle wave-vector $\mathbf{k}$ and $\mathcal{C}_{s}>0$ is a constant
whose particular form depends on $s$. The physical cases $s=2,\, 1$
correspond, respectively, to the nonrelativistic IFG with $C_{2}=\frac{\hbar^{2}}{2m}$ and
to the ultrarelativistic IFG for which $C_{1}=c\hbar,$ $c$ being the speed of
light. In this case $G_{d,s}$ takes the form $\mathcal{V}/[2^{d-1}\pi^{d/2}\Gamma(d/2)sC_{s}^{d/s}],$ with $\Gamma(\sigma)$ the gamma function and $\mathcal{V}=L^{d}$ the volume of the system. The second instance is based on the $d$-dimensional IFG trapped by an isotropic potential of the form $U(\mathbf{r})=U_{0}\left(r/r_{0}\right)^{\alpha}$, where $U_{0},$ $r_{0},$ are two constants that characterize the energy and length scales of the trap. This trapping potential leads, in the semi-classical approximation \cite{BagnatoPRA87}, to  $G_{d,s}=\frac{(2/s-1)\Gamma\left[d(1/s-1/2)\right]}{\Gamma(d/2)\Gamma[d/s]\hbar^{d}}
\left(\frac{mr_{0}^{2}}{2}\right)^{d/2}U_{0}^{d(1/2-1/s)}$ with $s^{-1}=1/2+\alpha^{-1}$. Notice that in the later case, one can immediately establish the thermodynamic equivalence between the IBG and the IFG, namely, $\alpha=2d/(2-d),$ implying that no such equivalence is possible in dimensions $d>2$ for positive $\alpha$. The equivalence does occur in two dimensions if $\alpha\rightarrow\infty,$ which corresponds to the infinite well potential  and in one dimension if $\alpha=2$, which corresponds to the harmonic potential.  

The thermodynamical properties of the ideal quantum gases are easily computed from the grand potential $\Omega(T,\mathcal{V},\mu)\equiv U-TS-\mu N$ \cite{PathriaBook,HuangBook}, where $U,$ and $S$ denote the internal energy, and entropy respectively. For the trapped gas, $\mathcal{V}$ denotes the appropriate thermodynamic variable that generalizes the \emph{volume} of a fluid in a rigid-walls container (see Ref. \cite{SandovalPRE2008} for the case of the three-dimensional harmonic trap), which in this paper is taken as $\mathcal{V}=\left(\frac{mr_{0}^{2}}{2}\right)^{d/2}U_{0}^{d(1/2-1/s)}$ which reduces to $\mathcal{V}=\omega^{-d}$ for the isotropic harmonic trap $U(\mathbf{r})=\hbar\omega\left(r/r_{0}\right)^{2}$ with $r_{0}=\left(2\hbar/m\omega\right)^{1/2}$. For a gas of noninteracting fermions, $\Omega(T,\mathcal{V},\mu)$ can be written in the thermodynamic limit, $N\rightarrow\infty$, $\mathcal{V}\rightarrow\infty$ with $N/\mathcal{V}=$ constant, \cite{footnote0} as 
\begin{multline}\label{GranPotential}
\Omega(T,\mathcal{V},\mu)=-k_{B}T\int_{0}^{\infty}d\varepsilon\, g(\varepsilon)\times\\
\ln\left[
\exp\{\beta(\varepsilon-\mu)\}f_{FD}(\varepsilon,T)\right],
\end{multline}
where $f_{FD}(\varepsilon,T)=\left\{\exp[\beta(\varepsilon-\mu)]+1\right\}^{-1}$ is the Fermi-Dirac distribution function that gives the average occupation of the single-particle energy state $\varepsilon$ at absolute temperature $T$. As usual, $\beta$ denotes the inverse of the product of $T$ and $k_{B}$ the Boltzmann's constant .

The average number of fermions $N(T,\mathcal{V},\mu)$ in the system is given by $-\left(\partial\Omega/\partial \mu\right)_{T,\mathcal{V}}$ \cite{PathriaBook} which
gives
\begin{align}  \label{NoEquation}
N/\mathcal{V}=-\overline{G}_{d,s}\, \Gamma(d/s)\, (k_{B}T)^{d/s}\, \hbox{Li}_{d/s}(-e^{\beta\mu}),
\end{align}
where $\hbox{Li}_{\sigma}(z)=\sum_{l=1}^{\infty}z^{l}/l^{\sigma}$ is the {\it polylogarithm} function of order $\sigma$ \cite{LewinBook} and $\overline{G}_{d,s}=G_{d,s}/\mathcal{V}$. Expression (\ref{NoEquation}) relates $N$ and $\mu$, and for fixed $N,$ the chemical potential is a function of the system temperature and volume. 
\begin{figure}[tbp]
\centering
\includegraphics[width=\columnwidth,clip=true]{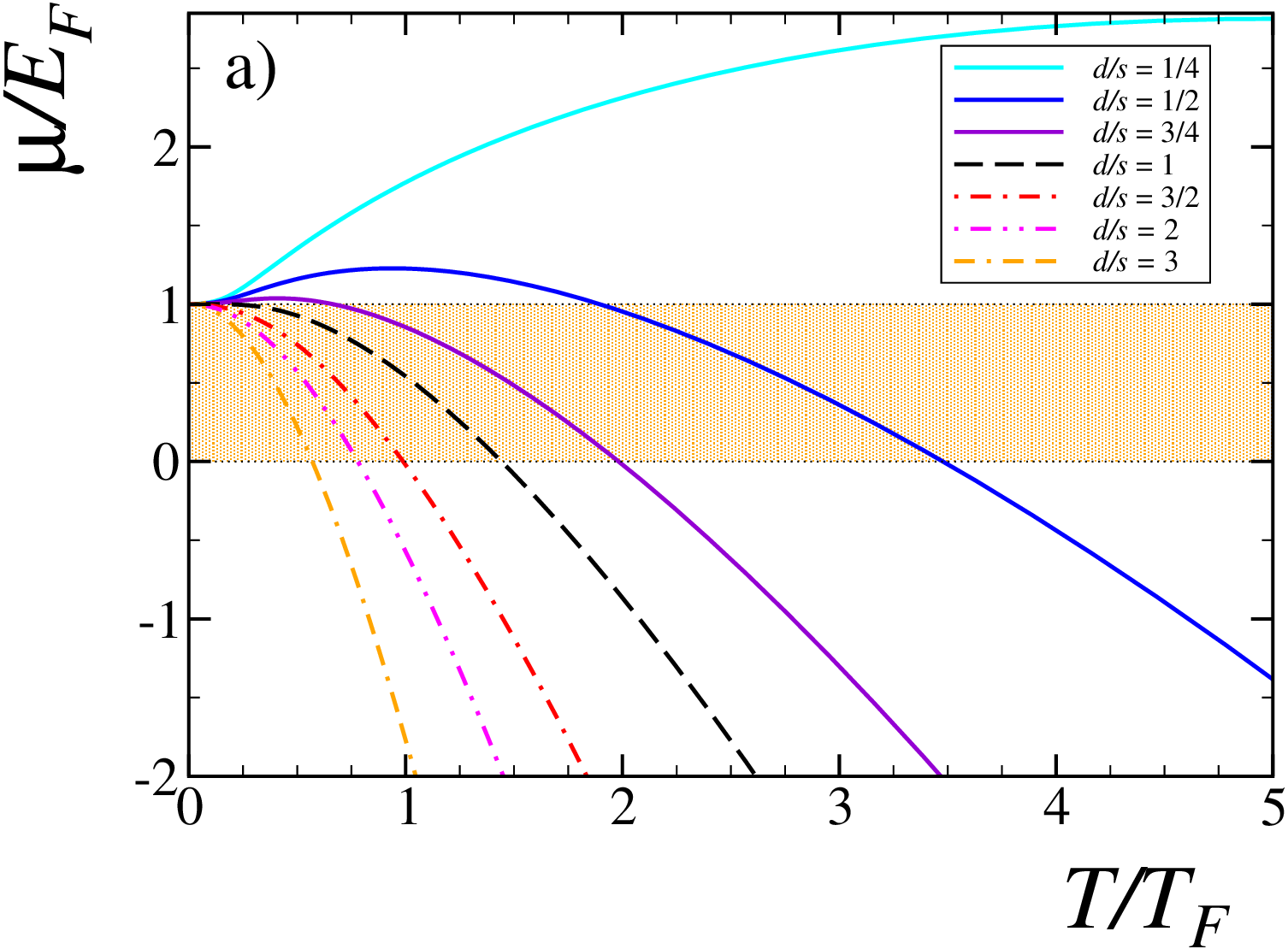}\\
\includegraphics[width=\columnwidth,clip=true]{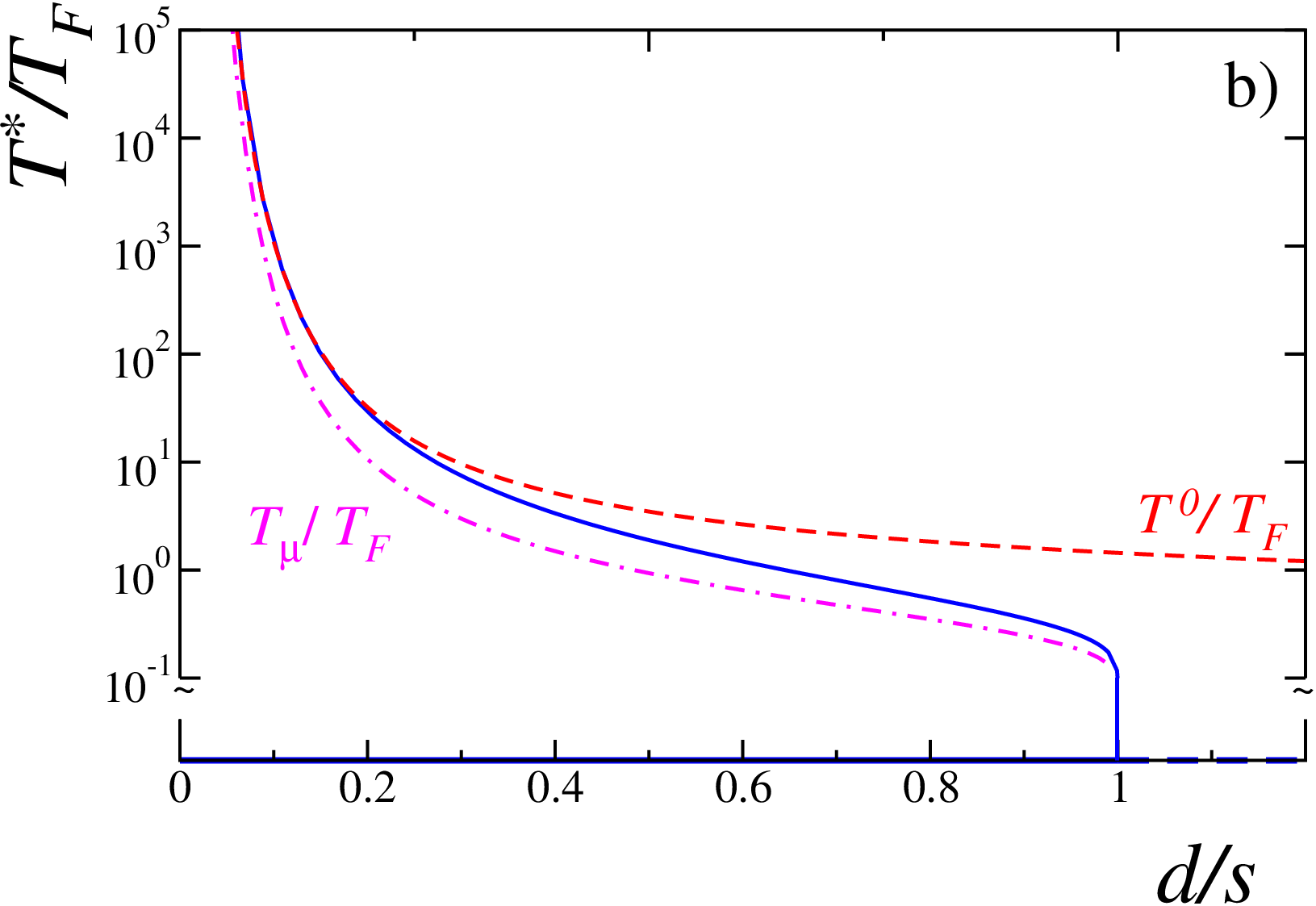}
\caption{(Color online) a) Dimensionless chemical potential $\mu/E_{F}$ as function of the dimensionless temperature $T/T_{F}$ for different values of $d/s.$ The crossings with the horizontal lines $\mu/E_{F}=1$ and $\mu/E_{F}=0$ marks the temperatures $T^{*}$ and $T^{0}$ respectively. b) The temperature $T^{*}$  as
function of $s/d$ (solid line), additionally the temperatures $T^{0}$ (dashed line) and $T_{\mu}$ (dash-dotted line) are included for comparison.}
\label{fig:1}
\end{figure}
The internal energy $U(T,\mathcal{V},\mu)$ per volume is given by 
\begin{align}\label{InternalEnergy}
U/\mathcal{V}=-\overline{G}_{d,s}\, \Gamma(d/s+1)\, (k_{B}T)^{d/s+1}\, \hbox{Li}_{d/s+1}(-e^{\beta\mu}),
\end{align}
while the entropy $S(T,\mathcal{V},\mu)=-\left(\partial\Omega/\partial T\right)_{\mathcal{V},\mu}$ per volume by 
\begin{multline}\label{Entropy}
S/\mathcal{V}=-k_{B}\overline{G}_{d,s}\, \Gamma(d/s)\, (k_{B}T)^{d/s}\left[\left(\frac{d}{s}+1\right)\right.\times\\
\left.\hbox{Li}_{d/s+1}(-e^{\beta\mu})-\frac{\mu}{k_{B}T}\hbox{Li}_{d/s}(-e^{\beta\mu})\right].
\end{multline}

In the top panel of Fig. \ref{fig:1} the temperature dependence of the ratio $\mu/E_{F}$ is shown for different values of the ratio $d/s$ and for $N/\mathcal{V}$ fixed, where $T_{F}$ denotes the Fermi temperature defined through the relation $E_{F}=k_{B}T_{F}$, where $E_{F}$ is explicitly given by $\left(\frac{d}{s \overline{G}_{d,s}}%
\right)^{s/d}(N/\mathcal{V})^{s/d}$ in $d$ dimensions. For $d/s<1$ the nonmonotonic dependence on temperature is clearly shown (the dashed line corresponds to the case $d/s=1/2,$ while $d/s=1/4$ is presented with the only purpose of making the effects of the system dimensionality more conspicuous). In the limit of high temperatures, $T\gg T_{F},$ the classical result $\mu\rightarrow k_{B}T\ln\left[\left(T/T_{F}\right)^{d/s}\Gamma(d/s+1)\right]$ is recovered.

As occurs for the 2D ideal gas in a box potential ($s=d=2$), the DOS is a constant whenever $s=d$, and the chemical potential has the well known analytical dependence on the temperature $\mu=E_{F}+k_{B}T\ln\left[1-e^{-T_{F}/T}\right].$ For $T\ll T_{F}$, the chemical potential lies below the Fermi energy by a negligible, exponentially small
correction. The low temperature behavior of $\mu$ for $d\ne s$ can be obtained approximately as a direct application of the Sommerfeld expansion for $T\ll T_{F}$ (see Ref. \cite{Ashcroft} pp. 45-46), namely
\begin{equation}  \label{sommerfeld}
\mu\simeq E_{F}\left[1-\frac{\pi^{2}}{6}\left(\frac{d}{s}-1\right)\left(%
\frac{T}{T_{F}}\right)^{2}\right]+\mathcal{O}\left([T/T_{F}]^{4}\right).
\end{equation}
The power-law dependence on $\varepsilon$ in expression (\ref{DOS}) is manifested itself in the last expression, where the ratio $d/s$ appears explicitly. Clearly, for $d/s<1,$ the chemical potential rises from the Fermi energy quadratically with $T$, and the non-monotonousness is a result of the fact that for large enough temperatures, $\mu(T)$ falls down with temperature to negative values close to those of the classical gas. As a consequence of this ``turning around'', $\mu(T)$ develops a maximum at temperature $T_{\mu}$ and equals $E_{F}$ at two distinct temperatures, at $T^{*}$ and 0, if $d/s<1$, and only at $T=0$ otherwise. Thus, the solution to the equation $\mu(T)=E_{F}$ as function of the parameter $d/s$, bifurcates at the critical value $d=s$ as is shown in the bottom panel of Fig. \ref{fig:1}. Note that for $s=2$ and $d=1$, $T^{*}$ is as large as $1.896\, T_{F}$ and diverges as $d/s\rightarrow0$. This can be shown straightforwardly from Eq. (\ref{NoEquation}) by putting $\mu=E_{F}$, since then, $T^{*}$ must satisfies the equation $1=\left[1+e^{-T_{F}/T^{*}}\right]^{-1}$ in that limit. 

In addition, the temperatures $T^{0}$ and $T_{\mu}$, that mark the change of sign of $\mu$ and its maximum, respectively, are also shown in the bottom panel of Fig. \ref{fig:1} (dashed line and dashed-dotted line). $T^{0}$ is determined from the equation $\mu(T^{0})=0,$ which explicitly gives 
\begin{equation}\label{T0}
T^{0}= \left[\Gamma(d/s+1)\, \zeta(d/s)(1-2^{1-d/s})\right]^{-s/d}T_{F},
\end{equation}
this expression gives the approximated values $3.48 \, T_{F},$ $1.44\, T_{F}$ and $0.989\, T_{F}$ for $d/s=1/2,\, 1,\, 3/2$ respectively. The temperature $T^{0}$ diverge as $\exp\{(s/d)\, \ln2\}$ as $d/s\rightarrow0$ and goes to zero as $[e/(d/s)]/\sqrt{2\pi d/s}^{s/d}$ as $d/s\gg1$, where $e$ is the Euler-Napier number.

It is clear from expression (\ref{sommerfeld}) that $d<s$ is required for the anomalous behavior of $\mu(T)$ to take place, however, physical positive integer dimensions less than three imposes severe restrictions on how fast the trapping potential must grow with the system size, i.e., on the values of the exponent $\alpha$. For fermions in a box-like trap ($s=2$) the anomaly will be observed if $d=1$, a case where the effects are conspicuously revealed even at large temperatures. This case indeed poses a challenge to trap designing, though, it could be realized experimentally by using the optical trap developed by Meyrath \textit{et al.} \cite{MeyrathPRA05}. In the typical experimental situation of harmonically trapped Fermi gases ($\alpha=2$ and therefore $s=1$) studied intensively, \cite{ButtsPRA97,GleisbergPRA00,Mueller04,VignoloPRA03} expression (\ref{sommerfeld}) tells us that the anomaly is not observed for any integer $d\ge1$. On the other hand, if one assumes $d=1$ as the minimum system dimensionality realizable experimentally (cigar shaped trapps), then one should go beyond harmonic trapping, \textit{i.e.}, one has to choose $\alpha>2$. 

\begin{figure}[tbp]
\includegraphics[width=\columnwidth,clip]{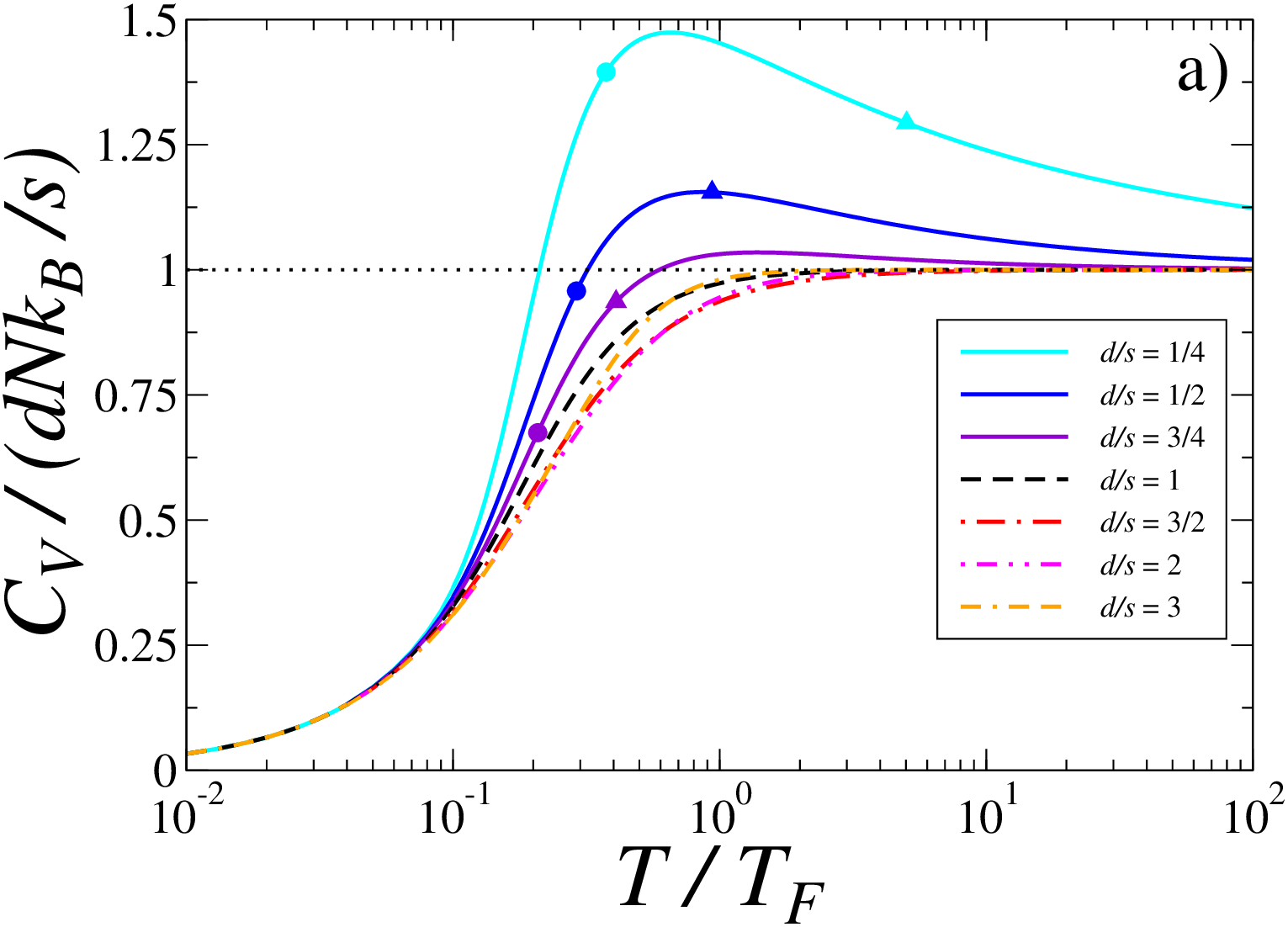}\\
\includegraphics[width=\columnwidth]{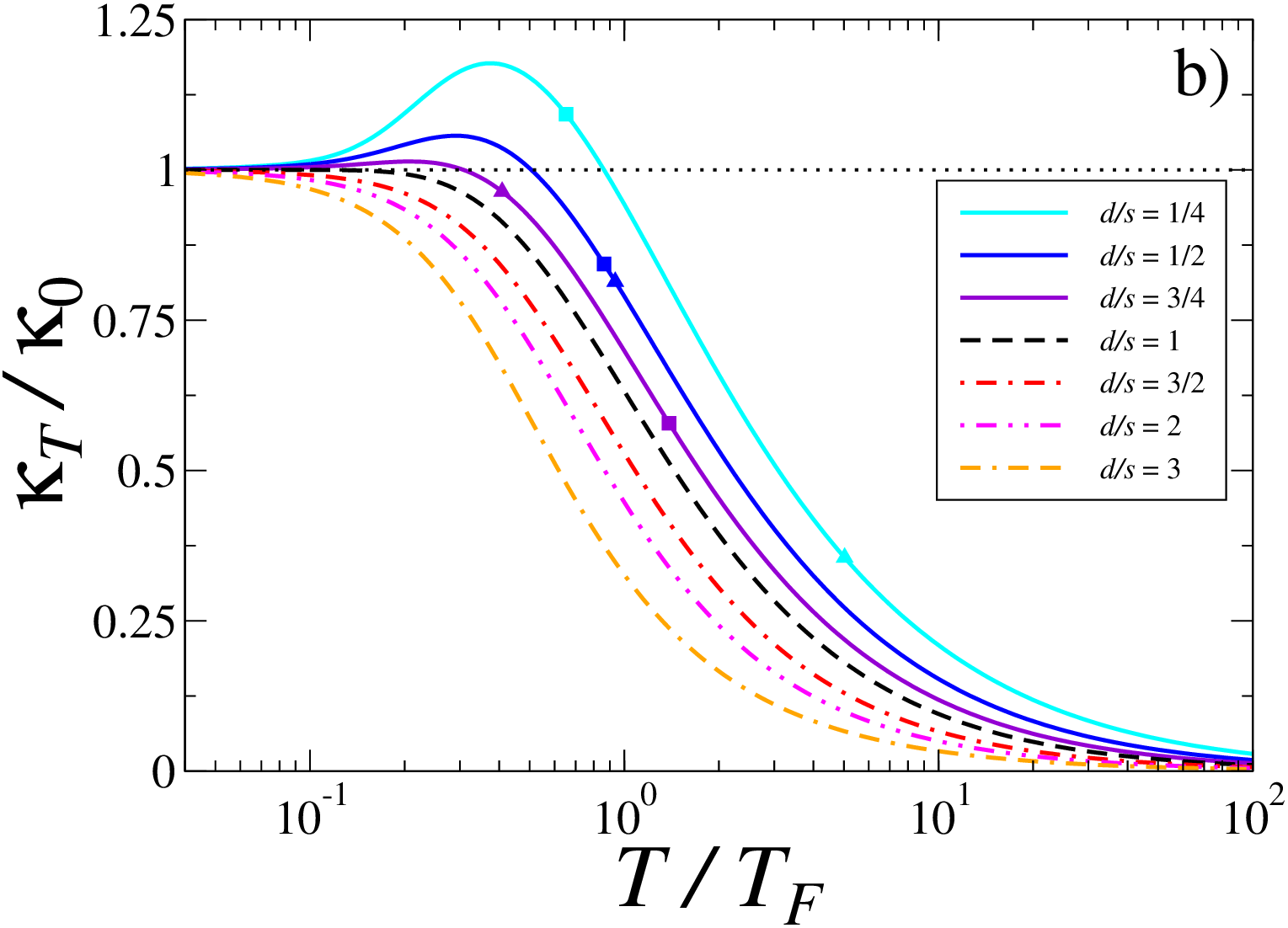}
\caption{(Color online) Normalized thermodynamic susceptibilities as function of the dimensionless temperature $T/T_{F}$ for different values of $d/s$, to say, 1/4, 1/2, 3/4, 1, 3/2, 2 and 3. a) Specific heat per particle at constant generalized volume, the circles mark the corresponding values of $C_{V}$ at $T_{\kappa_{T}}$ that is at the temperature at which $\kappa_{T}$ has a maximum values, analogously, the triangles do the same for at $T_{\mu}$ where $\mu$ has a maximum. b) Isothermal compressibility scaled with $\kappa_{0}= \left(\frac{d}{s}\right)^{2} \frac{\pi^{d}\Gamma(d/2)}{2\pi^{d/2}} sC_{s}^{^{d/s}} E_{F}^{-(d/s+1)}$, rhombus mark the corresponding values of $\kappa_{T}$ at $T_{C_{V}}$ that corresponds to the temperature at which $\kappa_{T}$ has a maximum values the triangles do the same as in a). Notice the nonmonotonic dependence on $T$ for $d/s<1$.}
\label{fig:2}
\end{figure}

The nonmonototicity of the chemical potential, just referred during the previous paragraphs, is revealed in the thermodynamic susceptibilities. In this work we focus on the specific heat at constant volume $C_{\mathcal{V}}=\left(\frac{\partial U}{\partial T}\right)_{\mathcal{V}}$ and the isothermal compressibility $\kappa_{T}=\frac{1}{n^{2}}\left(\frac{\partial n}{\partial\mu}\right)_{T}$, given by 
\begin{subequations}\label{susceptibilities}
\begin{multline}
\frac{C_{\mathcal{V}}}{Nk_{B}}=\frac{d}{s}\left(\frac{d}{s}+1\right)\frac{\text{Li}_{d/s+1}(-e^{\beta\mu})}{\text{Li}_{d/s}(-e^{\beta\mu})}\\
-\left(\frac{d}{s}\right)^{2}\frac{\text{Li}_{d/s}(-e^{\beta\mu})}{\text{Li}_{d/s-1}(-e^{\beta\mu})},\label{Cv}
\end{multline}
\begin{equation}
\kappa_{T}=\frac{\mathcal{V}}{Nk_{B}T}\frac{\text{Li}_{d/s-1}(-e^{\beta\mu})}{\text{Li}_{d/s}(-e^{\beta\mu})},\label{KappaT} 
\end{equation}
\end{subequations}
respectively. 

\begin{table}
\caption{\label{tabla:1} The temperatures $T^{0}$,  $T^{*}$, $T_{\mu}$, $T_{C_{V}}$ and $T_{\kappa_{T}}$ for which $\mu(T^{0})=0,$ $\mu(T^{*})=E_{F},$ $\mu(T_{\mu})$ is maximum, $C_{V}(T_{C_{V}})$ is maximum and $\kappa_{T}(T_{\kappa_{T}})$ is maximum, for three characteristic values of $d/s$, namely 1/4, 1/2 and 3/4 at which a non-monotic behavior is observed.}
\begin{ruledtabular}
\begin{tabular}{cccccc} 
$d/s$ & $T^{0}$ & $T^{*}$ & $T_{\mu}$ & $T_{C_{\mathcal{V}}}$ & $T_{\kappa_{T}}$ \\ 
\hline
0.25 & 15.6729 & 13.2260  & 5.0286 & 0.6532 & 0.3751 \\
0.5 & 3.4797 & 1.8960 & 0.9365 & 0.8632 & 0.2906 \\
0.75 & 1.9830 & 0.6666  & 0.4086 & 1.3893 & 0.2080 
\end{tabular}
\end{ruledtabular}
\end{table}
In Fig. \ref{fig:2} the dimensionless $C_{\mathcal{V}}(T)\, s/dNk_{B}$ (top panel) and $\kappa_{T}/\kappa_{0}$ (bottom panel) are shown as function of the dimensionless temperature $T/T_{F}$ for different values of $d/s$, clearly, for $d/s<1$, both quantities exhibit a non-monotonous dependence on $T$. The specific heat clearly exhibit the \emph{universal} linear dependence on $T$ in the low temperature regime and rises with temperature evidencing the effects of dimensionality. In the high temperature regime all the curves converge to the classical result $dNk_{B}/s$. Analogously, the isothermal compressibility exhibits the universal behavior in the low-temperature regime namely a finite value due to the degeneracy pressure. As temperature rises the effects of dimensionality are uncovered but are hidden again in the high temperature regime, where the classical dependence on temperature appears.

The nonmonotonic dependence with temperature of both thermodynamic susceptibilities is manifested as a global maximum at the temperatures $T_{C_{\mathcal{V}}}$ and $T_{\kappa_{T}}$, respectively (see solid lines in both panels of Fig. \ref{fig:2}). One would be tempted to propose that either of these temperatures would distinguish between two distinct behaviors of the IFG: one where the corresponding susceptibility behaves anomalously and other where it behaves standardly. Notice nevertheless, that such temperatures do not match between them nor with any of the temperature $T_{\mu}$ or $T^{*}$, as can be quantitatively appreciated in Table \ref{tabla:1} and in Fig. \ref{fig:2}, where solid triangles in both panels identify the values of the corresponding susceptibility evaluated at $T_{\mu}$, solid circles in panel a) indicate the values of $C_{\mathcal{V}}$ at $T_{\kappa_{T}}$ and analogously, solid squares in panel b) mark the value of $\kappa_{T}$ at $T_{C_{\mathcal{V}}}$ for $d/s=1/4,\, 1/2,$ and 3/4). Such discrepancy among all these temperatures makes difficult to consider them as points that mark the separation of two distinct thermodynamic behaviors.

For the signal value $d/s=1$, expressions in terms of elementary functions are possible (black-dashed lines in Fig. \ref{fig:2}), namely
\begin{subequations}
\begin{align}
 \frac{C_\mathcal{V}}{Nk_{B}}&=2\frac{\hbox{Li}_{2}(e^{\beta\mu})}{\ln(1+e^{\beta\mu})} - (1+e^{-\beta\mu})\ln(1+e^{\beta\mu})\\
 \kappa_T&= \frac{\mathcal{V}}{Nk_B T} \frac{e^{\beta\mu}}{(1+e^{\beta\mu})\ln(1+e^{\beta\mu})}
\end{align}
\end{subequations}
where we have used that the polylogarithm functions of order $0,1,2$ correspond to the elementary funtions $\hbox{Li}_{0}(z)=\ln(1+x)$, $\hbox{Li}_{1}(z)=z/1+z$ and $\hbox{Li}_{2}(z)=\int_{0}^{z}x\text{d}x/1+x$ respectively. For $d/s>1$, the variation with temperature of the thermodynamic susceptibilities is standard.

\section{\label{sect:3}Heuristic explanation of the nonmonotonic dependence of $\mu$ on $T$ for $d/s<1$}

\begin{figure}
 \includegraphics[width=\columnwidth,clip]{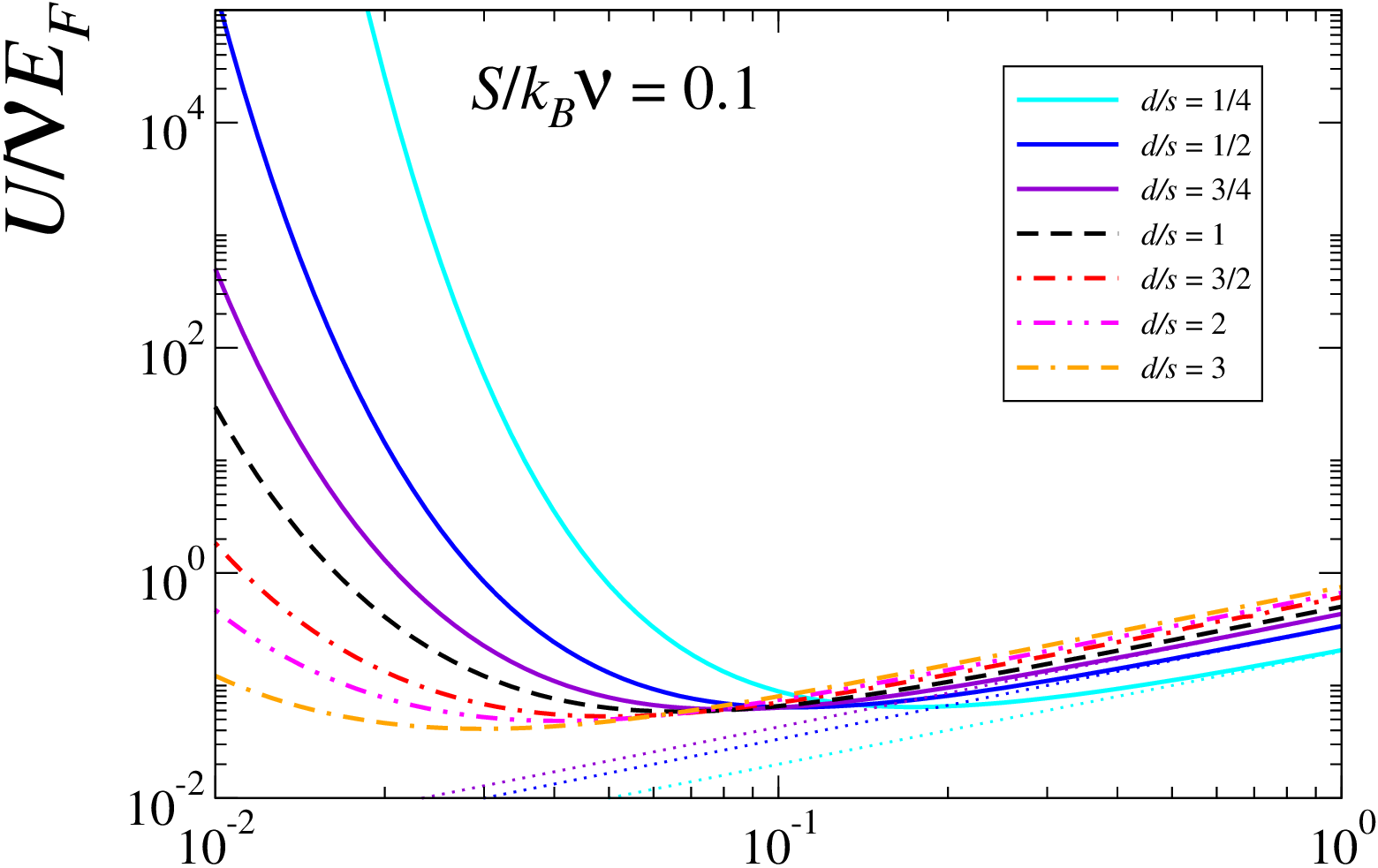}
 \includegraphics[width=\columnwidth,clip]{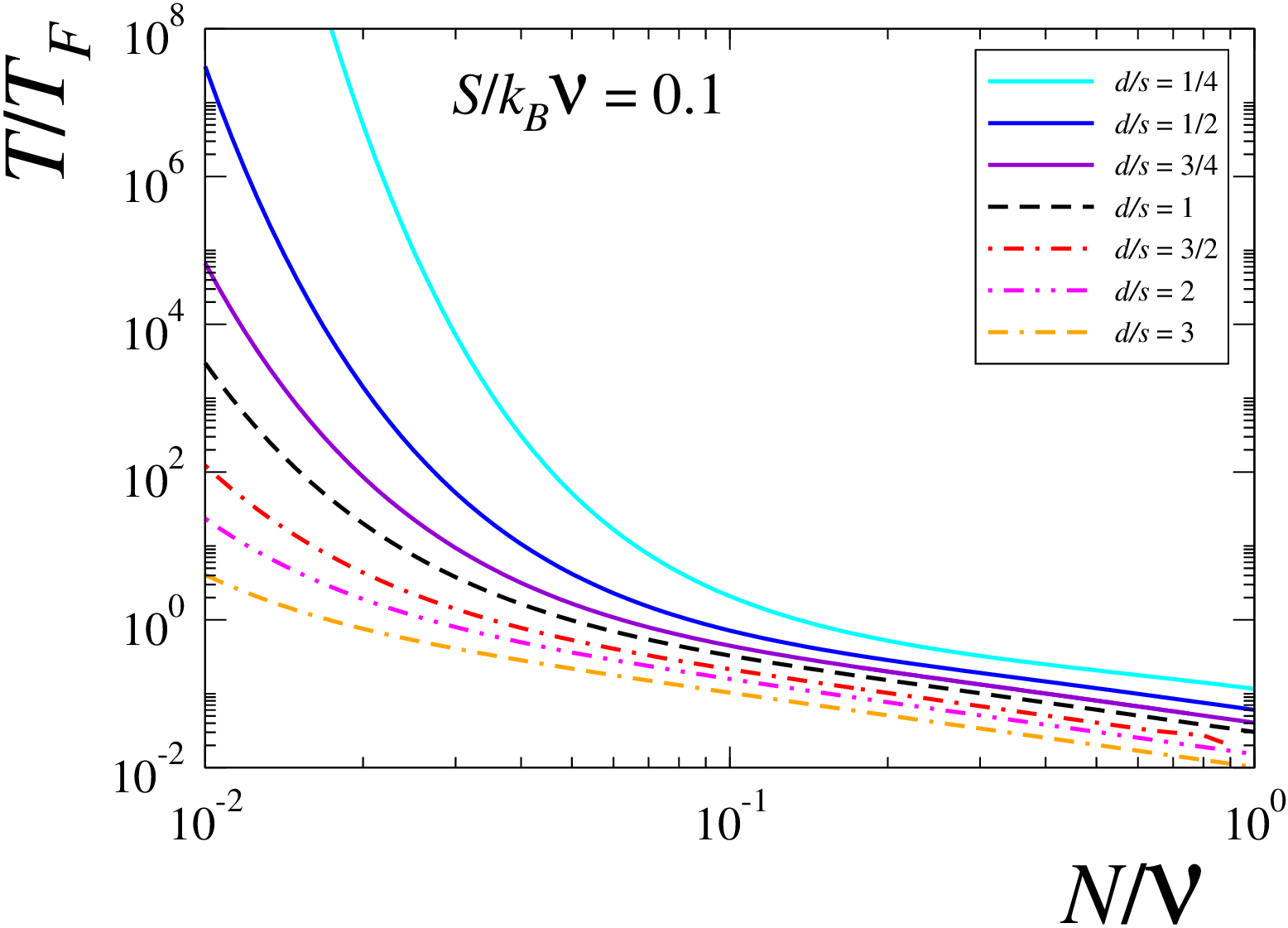}
 \caption{(Color online) The dependence of the scaled internal energy $U/\mathcal{V}E_{F}$ (top panel) and scaled temperature $T_/T_{F}$ (bottom panel) as function of the dimensionless particle density $N/\mathcal{V}$, where $\mathcal{V}$ denotes the systems volume scaled with an arbitrary volume $\mathcal{V}_{0}$.}
 \label{fig:3}
\end{figure}

The monotonic decreasing behavior of the chemical potential with temperature for $d/s\ge1$ is understood from the argument based on the fact that the internal energy $U$, diminishes from its zero temperature value $E_{F}$ after adiabatically adding a fermion at the small temperatures $T\ll T_{F}.$ Quoting Cook and Dickerson \cite{cook_ajp95}, the system cools by redistributing the particles into the available energy states in such a way that the particle added goes into ``\ldots a low lying, vacant single particle state, which will be a little below $E_{F}$''. This is a consequence, as we will show below, that in the three-dimensional case the change of the Helmholtz free energy is dominated by the change of entropy in the low temperature limit, however, the argument provided in \cite{cook_ajp95}, does not give neither the amount of the energy change involved in the process nor the change in temperature, making the nature of the argument just qualitative. In fact, the difficulty in quantifying those quantities arises from the use of the thermodynamic relation
\begin{equation}
\mu(S,\mathcal{V},N)=\left(\frac{\partial U}{\partial N}\right)_{S,\mathcal{V}}  \label{mu_U}
\end{equation}
which requires the knowledge of $U(S,\mathcal{V},N),$ rarely considered for analysis in the variables $S,\mathcal{V},N$. From \eqref{GranPotential} the functions $N=N(\mu,T,\mathcal{V}),$ Eq. \eqref{NoEquation}, and $S=S(\mu,T,\mathcal{V})$, Eq. \eqref{Entropy}, are obtained and solved in order to obtain $U(S,\mathcal{V},N)$. In Fig. \ref{fig:3} the internal energy at constant entropy is plotted as function of the particle density $N/\mathcal{V}$ for $S/k_{B}\mathcal{V}=0.1$ and for different values of the ratio $d/s$. The slope of the curves give the value of the chemical potential as given by expression \eqref{mu_U}. Also in the same Fig. \ref{fig:3}, but in the bottom panel, the temperature of the system, scaled with the Fermi temperature, as function of the particle density is shown for $S/k_{B}\mathcal{V}=0.1.$ Clearly, the systems cools regardless of the ratio $d/s,$ when adding particles to the system in an isentropic way. 

It is possible to obtain an expression for $\mu(S,\mathcal{V},N)$ from \eqref{mu_U} by the use of the asymptotic behavior of the Polylogarithm functions $-\hbox{Li}_{\sigma}(-z)\simeq\ln(z)^{\sigma}/\Gamma(\sigma+1)+(\pi^{2}/6)\ln(z)^{\sigma-2}/\Gamma(\sigma-1)+\ldots$, after some algebra we have expressly that in the degenerate regime
\begin{equation}
 \mu(S,\mathcal{V},N)\simeq E_{F}\left[1+\frac{3}{2\pi^{2}}\left(\frac{S}{k_{B}\mathcal{V}}\right)^{2}\left(\frac{\mathcal{V}}{N}\right)^{2}\frac{s}{d}\left(\frac{s}{d}-1\right)\right],
\end{equation}
where the nonmonotonic dependence on $T$ is evident when $d/s<1.$
On the other hand, for the sake of completeness we compute the system temperature as function of the particle density, in the degenerate limit, which is given by
\begin{equation}
 T(S,\mathcal{V},N)\simeq \frac{3}{\pi^{2}}\frac{s}{d}T_{F} \frac{S}{k_{B}\mathcal{V}}\frac{\mathcal{V}}{N}.
\end{equation}
and, as is shown in the lower panel of Fig. \ref{fig:3}, decrease as $(N/\mathcal{V})^{-1}$ .

How can we understand the rising of the chemical potential when $d/s<1$? Consider the number of particles that can be excited by the energy $k_{B}T\ll E_{F}$ from the $d$-dimensional Fermi sphere. This number is approximately given by $Nk_{B}T/E_{F}$ while the number of available states above the Fermi energy can be approximated by $g(E_{F})k_{B}T$. The quotient between both quantities is exactly $s/d.$ This simple and heuristic argument shows that there are more single-particle excited states than excitable particles for $d/s>1$, which is evident because of the monotonic increasing behavior of the DOS. In principle all the excited particles can be accommodated into the available states without violating Pauli's principle. The accommodation, however, is not arbitrary. The probability of occupation of the available states in thermal equilibrium must follow the Fermi-Dirac distribution and therefore just a fraction of the excitable fermions are excited into the interval $[E_{F}, E_{F}+k_{B}T]$ (in fact, the occupation probability for the states with energy larger than $\mu$ is smaller than $1/2$). For this case we can certainly apply the argument given by Cook \textit{et al.} in Ref. [\onlinecite{cook_ajp95}] to infer that when adding adiabatically an extra particle to the system, the internal energy will decrease from $E_{F}$.

In contrast, Pauli exclusion principle prohibits complete accommodation when $d/s<1,$ since in this case the DOS has a monotonic decreasing dependence on energy and, as consequence, the number of available excited states is reduced considerably in comparison with excitable number of particles. We may conclude that when adding a particle in an adiabatically way, the probability of occupying an energy state below $E_{F}$ is very small and therefore, it will occupy an energy state above $E_{F}$. 

\begin{figure}[tbp]
\centering
\includegraphics[width=\columnwidth]{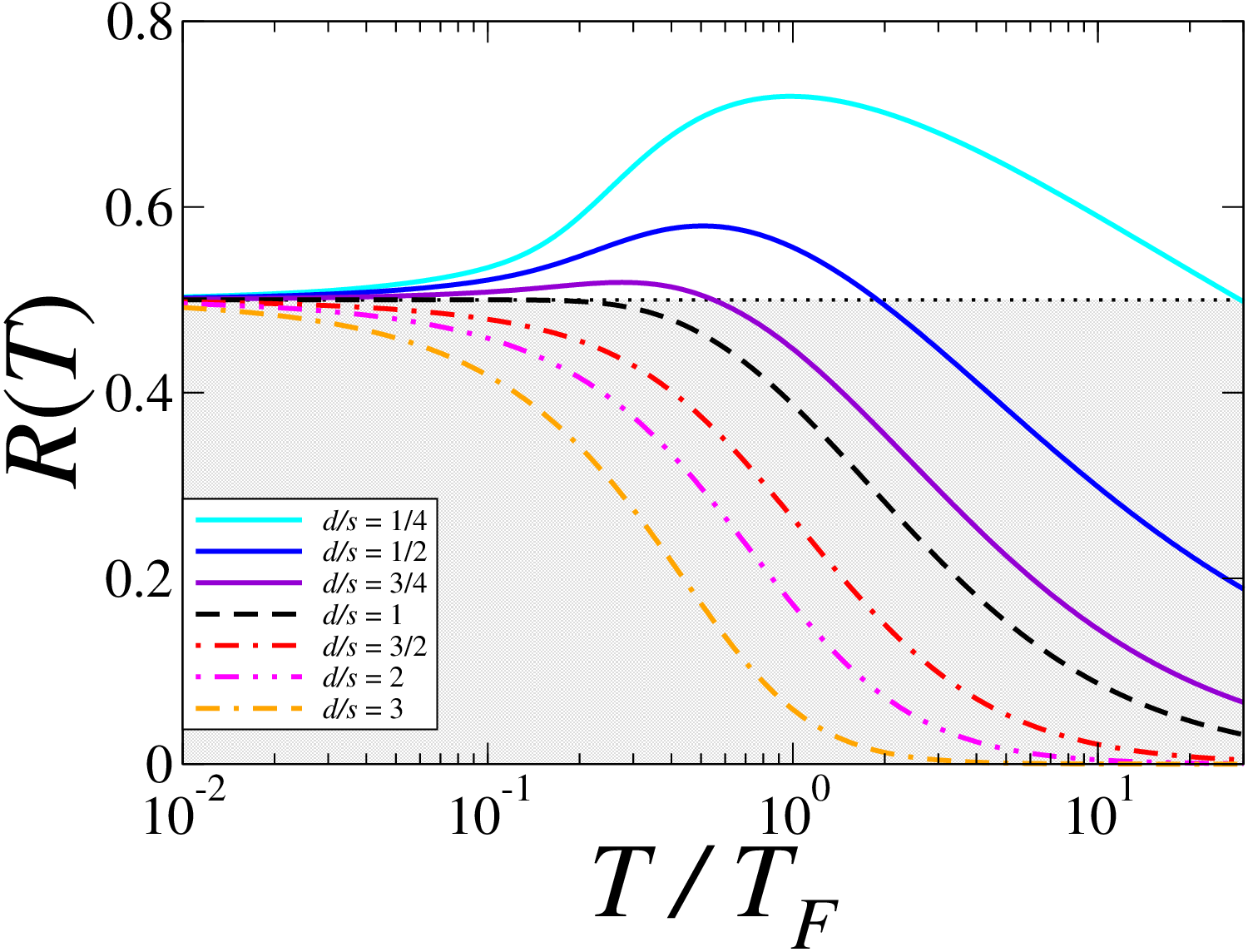}
\caption{(Color online) The ratio of the number of particles in the energy interval $\left[E_{F},E_{F}+\Delta\right]$ to the number of
available states in the same energy interval, $R(T)$, see Eq. \eqref{R}, as function of temperature for different values of $d/s.$ 
}
\label{fig:4}
\end{figure}

In order to quantitatively characterize the incomplete accommodation described above, we consider the ratio $R(T)$ of the number of particles in the energy interval $\left[E_{F},E_{F}+\Delta\right]$ to the number of
available states in the same energy interval, 
\begin{equation}\label{R}
R(T)=\frac{\int\limits_{E_{F}}^{E_{F}+\Delta}d\varepsilon\,
g(\varepsilon)f_{FD}(\varepsilon,T)}{\int\limits_{E_{F}}^{E_{F}+\Delta}d%
\varepsilon\, g(\varepsilon)},
\end{equation}
This quantity is shown in Fig. \ref{fig:4} as function of temperature with ratio $\Delta/E_{F}=0.001$, for different values of $d/s.$ 
The choice $\Delta\sim k_{B}T\ll E_{F}$ guarantees that a negligible number of particles occupy states out of the interval $[E_{F},E_{F}+\Delta]$. Under this condition we can approximate $R(T)$ by $f_{FD}(E_{F},T)$ and for temperatures $0<T\ll T_{F}$ we have that $R(T)\simeq\frac{1}{2}\left[1-\frac{\pi^{2}}{12}(\frac{d} {s}-1)\frac{T}{T_{F}}\right],$ therefore, the occupation probability of the energy states in $\left[%
E_{F},E_{F}+\Delta\right]$ is smaller than $1/2$ for $d/s>1$, greater than $1/2$ for $d/s<1$, and equal to  $1/2$ for $d=s$.

\section{\label{sect:4}The physical meaning of $T^{0}$}
\begin{figure}[h]
 \includegraphics[width=\columnwidth,clip]{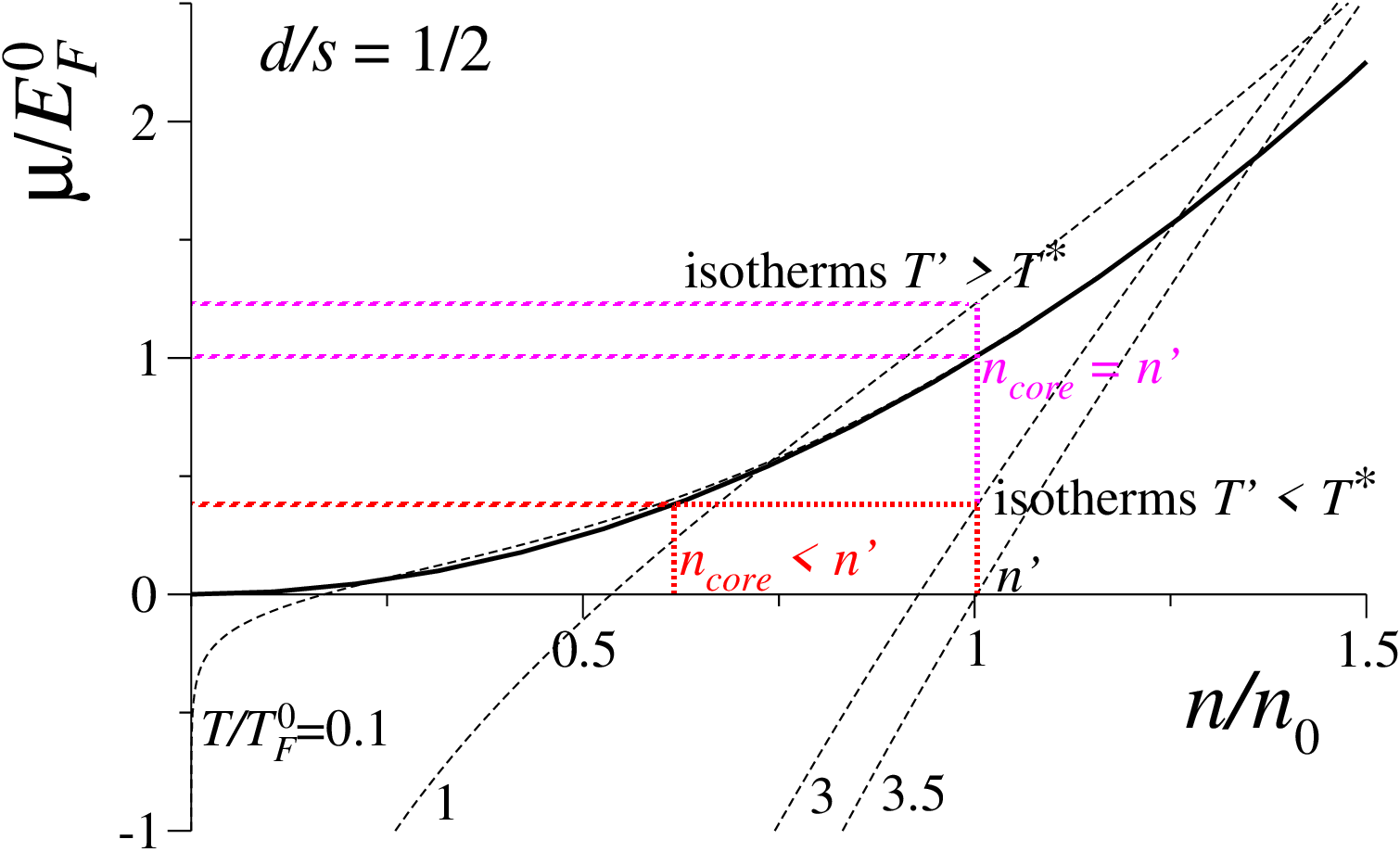}
 \caption{Isotherms in the plane $\mu$-$n$ for the case $d/s=1/2.$ Chemical potential and particle density are scaled with $E_{F}^{0}$ and $n_{0}$, respectively, which correspond to the Fermi energy of an IFG with an arbitrary particle density $n_{0}$. The thick line corresponds to the zero temperature isotherm, while thin-dashed lines label the isotherms with scaled temperatures $T/T_{F}^{0}=0.1,\, 1,\, 3,\,$ and 3.5. For isotherms lying in the region $\mu>0$ but below the zero temperature isotherm it is possible to find $n_{core}$.}
 \label{fig:5}
\end{figure}

A condensation-like phenomenon has been suggested to occur in the IFG in Refs. \cite{AnghelJPhysA03,Romero2011}, this can be understood as the formation of a ``core'' in momentum-space, reminiscent of the Fermi sea, that starts forming at $T^{0}$ and that grows up to form the Fermi sea as temperature is diminished to absolute zero. The number of particles in the core, $n_{core},$ is computed as follows \cite{Romero2011}: for a given value of the system density, lets say  $n^{\prime}$, and a temperature $T^{\prime}$, $n_{core}$ is found on the $\mu$-$n$ plane as the value of $n$ that corresponds to the intersection of the horizontal line $\mu(T^{\prime},n^{\prime})$ with the isotherm $\mu_{0}(n)=\mu(T=0,n)$ (thick line in Fig. \ref{fig:5} corresponds to $d/s=1/2$). Necessarily, such a process can not be performed at constant density implying an exchange of particles with and external reservoir in thermodynamic equilibrium with the system.  

\begin{figure}[h]
 \includegraphics[width=\columnwidth,clip]{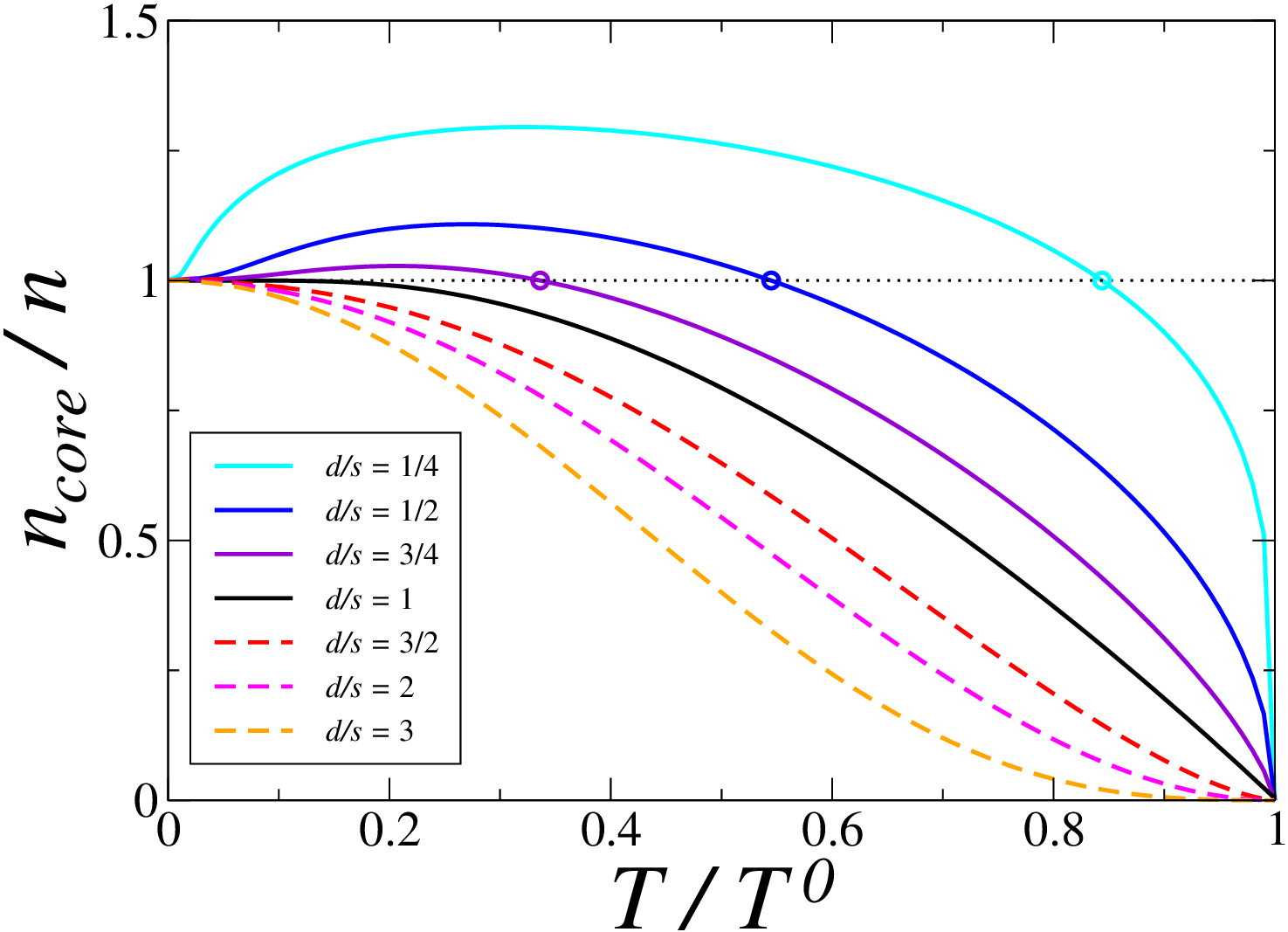}
 \caption{(Color online) Fraction of particles in the Fermi-sphere-like condensate as function of temperature scaled with $T^{0}$ for different values of $d/s$.}
 \label{fig:6}
\end{figure}

Is evident that no such intersection exist if $\mu(T^{\prime},n^{\prime})<0$, i.e. no interpretation of a core can be formulated in the non-degenerate regime, however, a solution $n_{core}\le n$ always exists for $\mu(T^{\prime},n^{\prime})>0$ and $d/s\ge1$, since the isotherm $\mu_{0}(n)$ is a concave function of the particle density, in other words, isotherms $\mu(T,n)$ of higher temperature are situated below $\mu_{0}(n)$ (see Ref. \cite{Romero2011} where the case $d/s=3/2$ is discussed). In contrast, for $d/s<1$, the zero temperature isotherm is a convex function of $n$ as shown in Fig. \ref{fig:5}, and two possibilities may happen: i) if $T^{\prime}<T^{*}$ the intersection occurs at $n_{core}<n^{\prime}$ (dark-broken lines in Fig. \ref{fig:5} for the case $d/s=1/2$); ii) on the contrary, if $T^{\prime}>T^{*}$ the intersection occurs at $n_{core}>n^{\prime}$ as shown explicitly in the same figure.
The dependence of the fraction $n_{core}/n$ on temperature is shown in Fig. \ref{fig:6} for different values of $d/s$, and is explicitly given by the expression
\begin{equation}
\frac{n_{core}}{n}=\left[\frac{\mu(T)}{E_{F}}\right]^{d/s}\quad \hbox{for } T\le T^{0}.
\end{equation}
Notice that the nonmonotonic dependence of $\mu(T)$ for $d/s<1$ makes $n_{core}/n$ to reach the value 1 at the temperature $T^{*}$ (see circles in Fig. \ref{fig:6}). 

\section{\label{sect:5}The argument energy-entropy and the meaning of $T^{*}$}
We now attempt to give a physical meaning to $\mu$ in the region where is larger than $E_{F}$, i.e. in the interval of temperatures $[0,T^{*}].$ For this purpose we compute $\mu(T,\mathcal{V},N)$ from the thermodynamic relation
\begin{equation}  \label{mu-canonical}
\mu(T,\mathcal{V},N)=\left(\frac{\partial F}{\partial N}\right)_{T,\mathcal{V}}
\end{equation}
where $F=F(T,\mathcal{V},N)$ stands for the Helmholtz free energy given by $F=-k_{B}T\ln
Z_{N,\mathcal{V}}(\beta)=U-TS$ with $Z_{N,\mathcal{V}}(\beta)=\sum_{E_{N,\mathcal{V}}}\exp\{-\beta E_{N,\mathcal{V}}\}$ the canonical partition function. The sum is made over the energies $E_{N,\mathcal{V}}$ of all possible configurations with exactly $N$ fermions in the volume $\mathcal{V}$. An advantage of expression \eqref{mu-canonical} over the use of the relation (\ref{mu_U}), is that at constant temperature and volume, the chemical potential measures the balance between the change of the internal energy and the heat exchanged when the number of particles in the system is varied from $N$ to $N+1$, making it suitable for the use of an \emph{energy-entropy} argument \cite{SimonJStatPhys1981}. Thus expression (\ref{mu-canonical}) provides a suitable operational definition, in the discrete case, of the chemical potential when only one particle is added isothermally to the system, namely \cite{Ashcroft,ShegelskiAJP2004,TobochnikAJP2005}
\begin{subequations}
 \begin{align}
 \mu(T,\mathcal{V},N)=&\Delta F\equiv F(N+1,T,\mathcal{V})-F(N,T,\mathcal{V}),\label{mucan2}\\
=&k_{B}T\ln\left[\frac{Z_{N,\mathcal{V}}(\beta)}{Z_{N+1,\mathcal{V}}(\beta)}\right].\label{mu3} 
 \end{align}
\end{subequations}
The rhs of expression \eqref{mucan2} can be explicitly written as $\Delta U-T\Delta S$, where $\Delta U=U(T,\mathcal{V},N+1)-U(T,\mathcal{V},N)$ and $\Delta S=S(T,\mathcal{V},N+1)-S(T,\mathcal{V},N),$ are the internal energy change of the system and the heat produced $T\Delta S$ when adding, isothermally, exactly one more fermion \cite{Note1}.

At zero temperature, the chemical potential is given by the change in internal energy only, whose value coincides with the Fermi energy of $N+1$ fermions, i.e. 
\begin{equation}
 \mu(\mathcal{V},N)=E_{F,N+1},
\end{equation}
where $E_{F,N}$ denotes the explicit dependence of the Fermi energy on the particle number. If this value is subtracted from \eqref{mucan2} we have that
\begin{equation}
 \Delta \mu = \Delta U^{\prime}-T\Delta S
\end{equation}
where $\Delta\mu=\mu(T,\mathcal{V},N)-E_{F,N+1}$ and $\Delta U^{\prime}=\Delta U-E_{F,N+1}$. In this way, if for a given temperature we have that $\Delta\mu\le0$, i.e., the chemical potential lies below the Fermi energy, then the relative change in the internal energy is smaller than the respective heat exchange by adding the particle. In other words, the effects of the addition of a particle to the system, in an isothermal way, are such that the entropic effects dominate over the energetic ones at that $T$. This argument accounts for the monotonic-decreasing behavior of $\mu$ with $T$, and is equivalent with argument given in Ref. \cite{cook_ajp95}. Further, if $\Delta\mu>0$ for a given $T$, then the energetic changes are the ones that dominate over the entropic ones, which give origin to the rise of $\mu$ above the Fermi energy as has been shown in the previous section. The temperature that separate both regimes coincides with $T^{*}$, which is different from zero when $d/s<1$. This suggest the possibility of interpreting $T^{*}$ as a critical temperature at which a phase transition occurs.

In order to show the validity of these ideas we first use expression (\ref{mu3}) to compute $\mu$ in two distinct one-dimensional systems each consisting of $N$ spinless fermions. One corresponds to the IFG trapped by a box-like potential ($d/s=1/2$), the other to the experimentally feasible system of and IFG trapped by a harmonic trap ($d/s=1$). We show that for the former case, $\mu$ rises above $E_{F,N+1}$ and eventually return to its decreasing behavior as the system temperature is increased from zero. For the later, we show that the $\mu<E_{F,N+1}$ for all $T>0.$

For exactly $N$ non-interacting fermions, the partition function satisfy the recursive relation \cite{BorrmannJChemPhys93,PrattPRL00}
\begin{equation}\label{Zrecursive}
Z_{N}(\beta)=\frac{1}{N}\sum_{n=1}^{N}(-1)^{n+1}Z_{1}(n\beta)Z_{N-n}(\beta),
\end{equation}
where $Z_{1}(\beta)=\sum_{\ve}\exp\{-\beta\ve\}$ is the single-particle partition function with $\varepsilon$ the single-particle energy spectrum and $Z_{0}(\beta)\equiv1$.

Expression (\ref{Zrecursive}) can be reduced to the calculation of $Z_{1}(m\beta)$, with $m$ a positive integer, by noting that $Z_{N}(\beta)$ can be written as a sum of the product over the distinct parts of all the partitions $\left\{(\lambda_{1},\lambda_{2},\ldots,\lambda_{r})\right\}$ of $N$ (a partition is defined as a nonincreasing sequence of positive integers $\lambda_{1},\lambda_{2},\ldots,\lambda_{r}$ such that $\sum_{i=1}^{r}\eta_{i}\lambda_{i}=N,$ where $\eta_{i}$ denotes the multiplicity of the part $\lambda_{i}$ in a given partition (for instance the partition $2+2+2+1$ of 7, the part 2 has multiplicity 3, see Ref. [\onlinecite{AndrewsBook}] pp.1), thus
\begin{equation}\label{Zpartition}
Z_{N}(\beta)= (-1)^{N}\sum_{\{(\lambda_{1},\lambda_{2},\ldots,\lambda_{r})\}}\prod_{m=1}^{r}\frac{(-1)^{\eta_{m}}}{\lambda_{m}^{\eta_{m}}\, \eta_{m}!}\left[Z_{1}(\lambda_{m}\beta)\right]^{\eta_{m}},
\end{equation}
The first four terms can be checked straightforwardly and are shown in Table \ref{Tabla:1}. 

\begin{table}[htb]
\begin{ruledtabular}
\caption{The firts 4 expressions for the canonical partition function are shown}
\begin{tabular}{cc}
$N$ & $Z_{N}(\beta)$\\
\hline
 1 & $Z_{1}(\beta)$\\
 2 & $\frac{1}{2}Z_{1}^{2}(\beta)-\frac{1}{2}Z_{1}(2\beta)$\\
 3 & $\frac{1}{6}Z_{1}^{3}(\beta)-\frac{1}{2}Z_{1}(\beta)Z_{1}(2\beta)+\frac{1}{3}Z_{1}(3\beta)$\\
 4 & $\frac{1}{24}Z_{1}^{4}(\beta)-\frac{1}{4}Z_{1}^{2}(\beta)Z_{1}(2\beta)+\frac{1}{3}Z_{1}(\beta)Z_{1}(3\beta)+$\\
   & $\hspace{5cm}\frac{1}{8}Z_{1}^{2}(2\beta)-\frac{1}{4}Z_{1}(4\beta)$\\
\end{tabular}
\label{Tabla:1}
\end{ruledtabular}
\end{table}

\begin{figure}[h]
\includegraphics[width=\columnwidth]{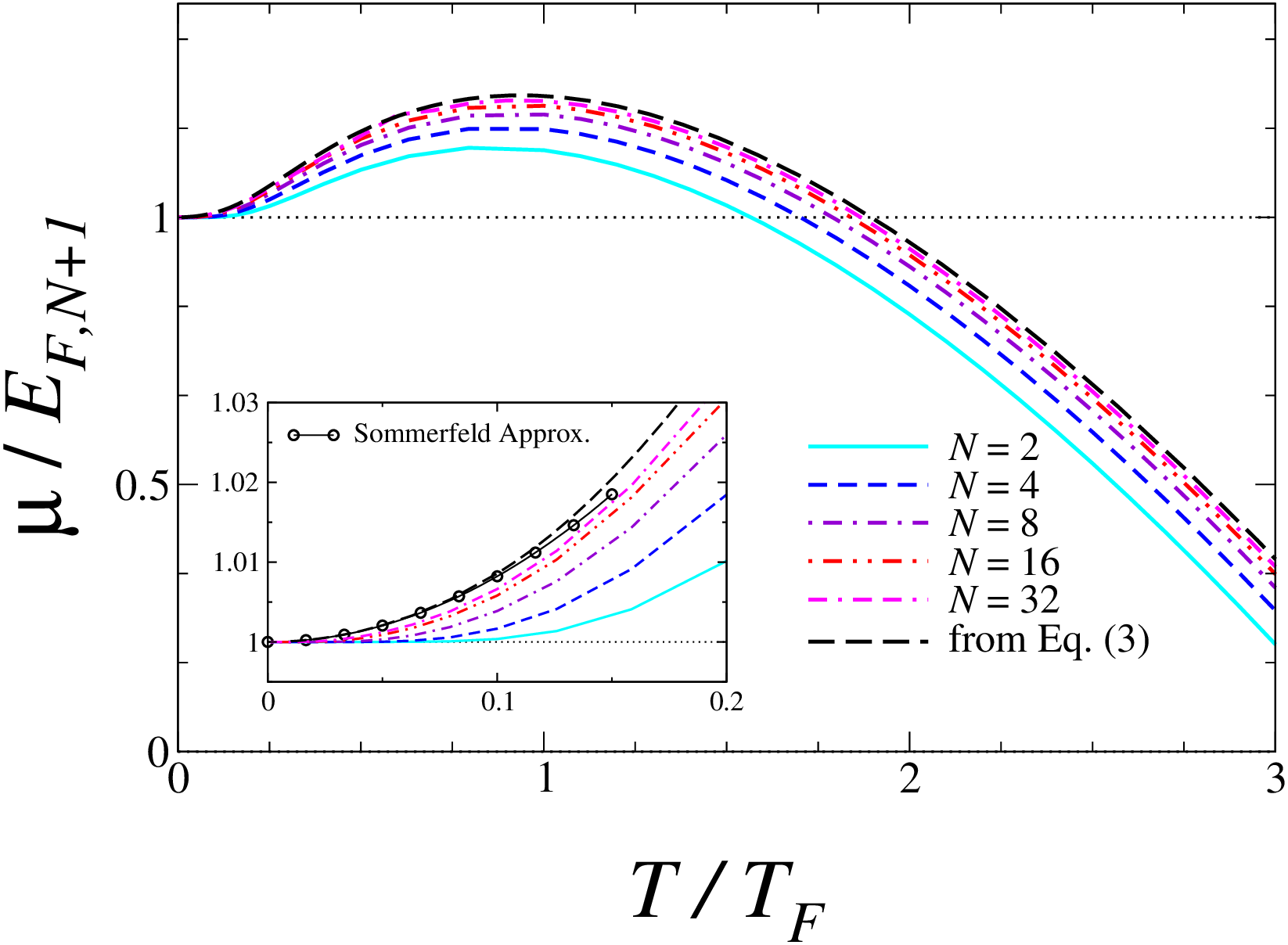}
\caption{(Color online) The chemical potential, Eq. \eqref{mu3}, scaled with $E_{F,N+1}$ as function of the dimensionless temperature $T/T_{F}$, with $T_{F}=E_{F,N+1}/k_{B}$, for the number of particles $N=2,\, 4,\, 8,\, 16,\,$ and $32$. Note that the results with 32 particles (double-dashed-dotted line) are close to the grand canonical ensemble result (black line in long dashes) computed from \eqref{NoEquation}. A comparison with the Sommerfeld approximation for low temperatures is shown in the inset.}
\label{fig:7}
\end{figure}

Computationally, evaluation of expression (\ref{Zpartition}) is faster than evaluating expression (\ref{Zrecursive}) since recursion is avoided and only the algorithm for computing the unrestricted partitions  of the integer $N$ is needed. Such algorithm forms part of the MATHEMATICA software package distribution. The computation time and memory requirements grow with number of partitions $p(N)$ of $N$, which grows asymptotically as $\exp\{\sqrt{n}\}$ thus limiting computation to $N\sim10.$ Suprisingly, the calculation exhibits a fast convergence to the well-known result obtained from \eqref{NoEquation} for 64 particles (see Fig. \ref{fig:7} for the box-like trap).

For the box potential in one dimension, the single-particle partition function is given in terms of the Jacobi theta function $\vartheta_{3}(u,q)=1+2\sum_{n=1}^{\infty}q^{n^{2}}\cos(2n        *u)$ as $Z_{1}(\beta)=\frac{1}{2}\left[\vartheta_{3}(0,e^{-\beta\varepsilon_{0}})-1\right]$ where the energy scale $\ve_{0}$ in the argument of the exponential is $\frac{\hbar^{2}\pi^{2}}{2mL^{2}}.$ For few particles, the chemical potential does not rise as $(T/T_{F})^{2}$ as is expected from the grand-canonical-ensemble result, but it grows much more slower as is shown in the inset of Fig. \ref{fig:7}. This is consequence of the low-temperature behavior of the partition function, which satisfies that $Z_{N}/Z_{N+1}\sim e^{\beta E_{F,N+1}}\left[1+e^{-\beta(2N+1)\varepsilon_{0}}\right]$ for $T/T_{F}\ll1,$ leading thus to $\Delta\mu\sim k_{B}Te^{-\beta(2N+1)\varepsilon_{0}}.$   

For the harmonic potential in dimension one, an exact analytical expression for $Z_{N}(\beta)$ is known \cite{TranPRE01,TranAnnPhys04}, namely 
\begin{equation}
Z_{N}(\beta)= \exp\left[-N^{2}\beta\frac{\hbar\omega}{2}\right]\prod_{j=1}^{N}\left[1-\exp(-\beta\hbar\omega\, j)\right]^{-1}.
\end{equation}
A direct application of Eq. (\ref{mu3}) leads to
\begin{equation}\label{mu1dHO}
\mu=E_{F,N+1}+k_{B}T\ln\left[1-\exp(-\beta E_{F,N+1})\exp(-\beta\hbar\omega/2)\right].
\end{equation}
Clearly (\ref{mu1dHO}) is a monotonically decreasing function of $T$ agreeing with the grand canonical ensemble result, expression $\mu=E_{F}+k_{B}T\ln\left[1-e^{-T_{F}/T}\right]$ is recovered in the limit $N\rightarrow\infty$.

\section{\label{sect:6}Conclusions and Final Remarks}
In this paper we have presented a discussion on the meaning of the nonmonotonic dependence on temperature of the thermodynamics properties of low dimensional, trapped, IFGs, with focus on the chemical potential (a similar behavior has been predicted for weakly repulsively interacting bose gases \cite{ZhangPRA09} in that a hard core Bose gas behaves, at least qualitatively, as an ideal Fermi gas). The parameter used to characterize the trapping and dimensionality $d$ of the system is merely $d/s$, that explicitly appears in the single-particle density of states \eqref{DOS}. Thus low dimensional trapped systems are characterized by values of $d/s<1.$ In this range of values, the chemical potential, the specific heat at constant volume and the isothermal compressibility, exhibit a nonmonotonic dependence on temperature which have been characterized by the temperatures $T_{\mu}$, $T_{C_{\mathcal{V}}}$, $T_{\kappa_{T}}$ \cite{GretherEPJD2003} respectively. We also have computed $T^{0}$ as function of $d/s$ \cite{AnghelJPhysA03,Romero2011}, and introduced a new characteristic temperature $T^{*}\le T^{0}$, which corresponds to the nonzero value of the temperature at which $\mu(T^{*},\mathcal{V},N)=E_{F}.$

We found that $T^{*}$ marks the temperature at which the particle density of a Fermi-like core $n_{core}$, that starts forming $T^{0}$, saturates at the value of the total particle density of the system $n$. This suggest that $T^{*}$ can be considered as the relevant temperature of the isotropically trapped IFG, as is supported by the energy-entropy-like argument presented in Sect. \ref{sect:5}. The region in the $\mu$-$T$ plane, for which $\mu>E_{F}$ for $T\le T^{*}$, represents the set of thermodynamic states for which the change in the Helmholtz free energy, when increasing the particle density of the system, is dominated by the changes of the internal energy and would correspond to an ordered phase. In the complementary region for which $T\ge T^{*}$, the thermodynamic states are characterized by changes in $F(T,\mathcal{V},N)$ dominated by heat exchange by changing entropy, and can be considered as a ``disordered phase''. Though, heuristic energy-entropy arguments have been used to uncover the possibility of a phase transition \cite{SimonJStatPhys1981}, we want to emphasize that we are not claiming the existence of a phase transition in the IFG, on the basis that thermodynamic quantities do not show a singular behavior of the thermodynamic susceptibilities at $T^{*}.$

Though the chemical potential is not directly measured in current experiments, development on imaging techniques of ultracold gases \cite{HoNaturePhys2010,SannerPRL2010,MullerPRL2010,NascimbeneNature2010} has open the possibility to experimentalists to measure the local particle-density \emph{in situ} and from the data to extract $\mu$ and $T$.

\begin{acknowledgments}
I want to express my gratitude to Mauricio Fortes and Miguel Angel Sol\'is for valuable discussions, and to Omar Pi\~na P\'erez for helping in generating figure 2. Support is acknowledge to UNAM-DGAPA PAPIIT-IN113114. 
\end{acknowledgments}


\begin{thebibliography}{99}
\bibitem{footnote00} Indeed, the precise argument presented in Ref. \onlinecite{cook_ajp95} is only valid for the free IFG in three dimensions.
\end{thebibliography}
\end{document}